\documentclass[12pt]{article}
\usepackage[dvips]{epsfig}
\title{
Exact Results on $\mathbf{{\cal O}(\alpha)}$ Corrections to the Single Hard
Bremsstrahlung Process in Low Angle Bhabha Scattering in the SLC/LEP Energy
Regime\thanks{Work supported in part by the US DoE, contract 
DE-FG05-91ER40627, and by the Polish Government, grant KBN 2P30225206.}
}
\author{
S.\ Jadach\thanks{Permanent address: Institute of Nuclear Physics,
          ul. Kawiory 26a, PL 30-059 Cracow, Poland},
          M. Melles, B.F.L.\ Ward, and S.\ A.\ Yost\\
{\normalsize\em Department of Physics and Astronomy, 
          University of Tennessee,}\\
      {\normalsize\em Knoxville, Tennessee 37996-1200, USA}
} 
\date{\normalsize January 1995\\[1 em] \normalsize UTHEP-95-0101}

\begin{document}
\maketitle
\baselineskip=0.6cm
\begin{abstract}
\baselineskip=0.6cm
We present the exact ${\cal O}(\alpha)$ correction to the process $e^+ e^-
\rightarrow e^+ e^- + \gamma$ in the low angle luminosity regime at
SLC/LEP energies.  We give explicit formulas for the completely
differential cross section.  As an important application, we illustrate
the size of the respective corrections of ${\cal O}(\alpha^2)$
to the SLC/LEP luminosity cross section.
We show explicitly that our results
have the correct infrared limit, as a cross-check.  Some comments are
made about the implementation of our results in the framework of a
Monte Carlo event generator.  This latter implementation will appear
elsewhere.
\end{abstract}
\newpage


\section{Introduction}

Recently, new luminometers at LEP\cite{lumin:1995} have made
measurements of the luminosity process $e^+ e^- \rightarrow e^+ e^- +
n(\gamma)$ at the experimental precision tags below $.1\%$.  This
should be compared with the prediction by two of us (S.J.  and
B.F.L.W.), with E.\ Richter-W\c as\ and Z.\ W\c as, of these
processes at the $.25\%$ precision tag in Ref.\ \cite{th6118:1991}
using the YFS Monte Carlo event generator BHLUMI2.00.  Recently, using
version 4.0 of BHLUMI, the authors in Ref.~\cite{th6118:1991} have
reported precision tags $\sim .1\%$ for the ALEPH SICAL detector's
asymmetric acceptance theoretical cross section
prediction~\cite{glasgow,th-9538:1995} and .16\% precision for the
general ALEPH SICAL acceptance.  The four of us, with W.  P\l aczek, E.
Richter-W\c as, M.  Skrzypek and Z.  W\c as~\cite{gatlin:1994}, have
reported the entirely equivalent conservative theoretical precision
tag $\sim .15\%$ for the bremsstrahlung correction for the general
acceptance of the ALEPH SICAL detector.  These results are currently
being extended to the other new LEP luminometers\cite{semi-paper}.  It
is clear that the precision tag on the theoretical prediction of
$\sigma_{{\cal L}}$, the SLC/LEP luminosity cross section, urgently
needs further improvement over the $\sim .15\%$ level if the
theoretical uncertainties are going to be reduced to the required
level of one half of $\Delta\sigma_{{\cal L}}^{\exp}$, the respective 
experimental uncertainty, so that they do not obscure the comparison between 
theory and experiment in the high precision $Z^0$ physics tests of the 
Standard Model of the electroweak interaction.

From Table 2 in Ref.\ \cite{th6118:1991} and from Table 1 in 
Ref.\ \cite{glasgow}, we see that the subleading part of the
${\cal O}(\alpha^2)$ pure bremsstrahlung correction to
$\sigma_{{\cal L}}$ remains, at this writing, a dominant part of the
outstanding theoretical uncertainty, where it contributes directly to
the physical precision and indirectly to the technical precision both
at a level $\sim .1\%$ itself.  Exact results on the double
bremsstrahlung process itself have been derived in Refs.\ \cite{zhang}
and \cite{twopho} and, in fact, the entire leading log part of the
${\cal O}(\alpha^2)$ correction to $\sigma_{{\cal L}}$ has recently
\cite{lep1-94a} been incorporated into BHLUMI4.0. What remains to be 
done rigorously then is to compute the remaining exact subleading part 
of the bremsstrahlung correction to $\sigma_{{\cal L}}$ and incorporate 
it into our YFS Monte Carlo event generator BHLUMI4.xx to check precisely 
the size of this effect in $\sigma_{{\cal L}}$.  

In this paper, we present the exact results
which are necessary to evaluate this remaining 
unquantified sub-leading part of the bremsstrahlung correction to
$\sigma_{{\cal L}}$ with particular emphasis on the acceptances of the 
new LEP high precision luminometers.  While partial results and estimates 
on this sub-leading correction to $\sigma_{{\cal L}}$ have appeared 
elsewhere\cite{kuraev-gatlinburg:1994,partial}, our results are the first 
ever, fully differential exact results of their type.  Their detailed 
implementation into BHLUMI4.xx will appear elsewhere\cite{semi-paper}.

More precisely, the lone outstanding contribution from bremsstrahlung
processes to $\sigma_{{\cal L}}$ which dominates the physical
precision part of the error in BHLUMI4.0 from ${\cal O}(\alpha^2)$
bremsstrahlung is the sub-leading part of the virtual correction to
the single bremsstrahlung process.  Thus, it is this latter process
which we shall compute exactly in what follows in a fully differential
manner, as is needed for Monte Carlo event generator applications.  We
repeat --- such a completely differential, exact ${\cal O}(\alpha^2)$
single bremsstrahlung calculation has not appeared elsewhere.  (See,
however, Refs.\ \cite{kuraev-gatlinburg:1994,partial} for various
levels of partial results.)

Our work is organized as follows.  In Sect.\ \ref{sec:prelim}, we set
our kinematic and notational conventions.  In Sect.\ \ref{sec:calc},
we analyze the processes of interest to us using the algebraic program
FORM \cite{FORM}.  In Sect.\ \ref{sec:results}, we present numerical
results which illustrate checks on our work
in the SLC/LEP luminosity regime.  Sect.\ \ref{sec:final} contains our
summary remarks. The Appendices contain some technical details.


\section{Preliminaries}
\label{sec:prelim}

In this section we set our kinematical notation and calculational
conventions.  We begin with the kinematics.

\begin{figure}
\begin{center}
\epsfig{file=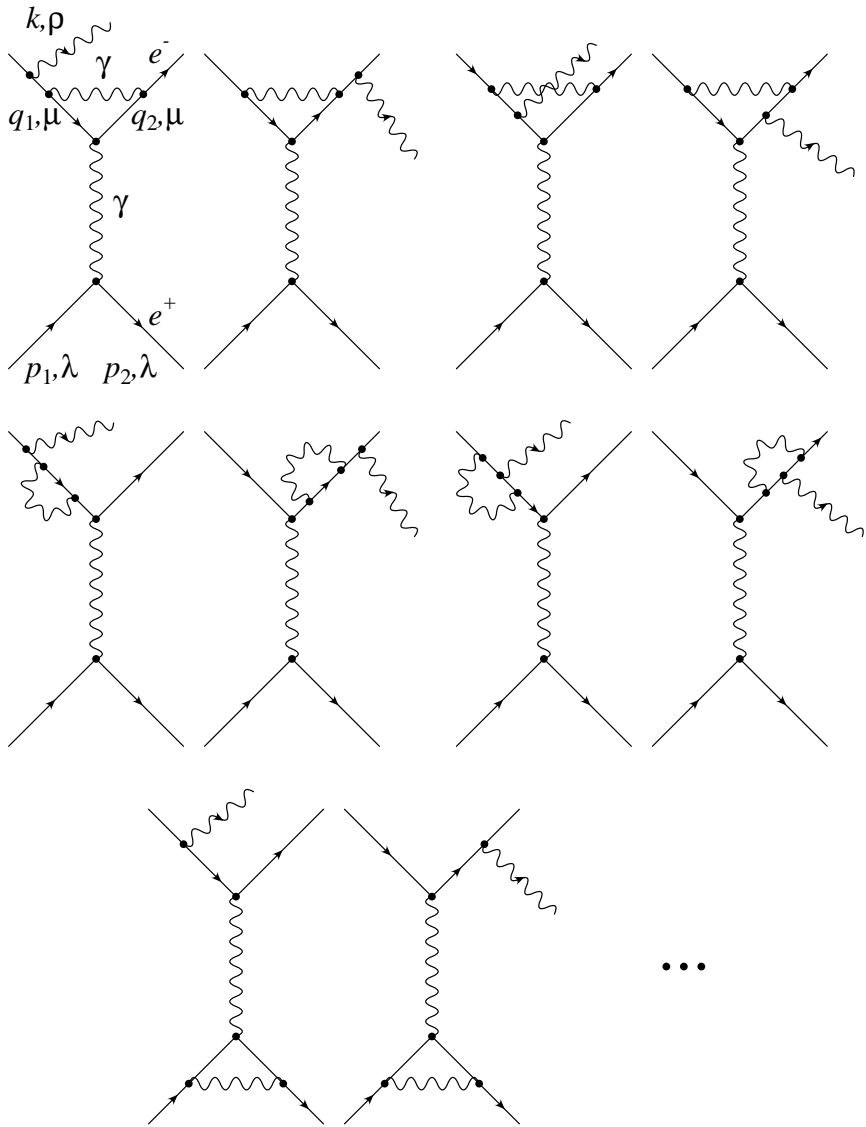,height=15cm}
\end{center}
\caption{${\cal O}(\alpha^2)$ single bremsstrahlung correction in
$e^+ e^- \rightarrow e^+ e^-$ at low angles. Only electron line emission
graphs are shown.}
\label{fig:process}
\end{figure}

The process under discussion is illustrated in Fig.\ \ref{fig:process}.  
We consider the one-loop corrections to the
process $e^+(p_1) + e^-(q_1) \rightarrow e^+(p_2) + e^-(q_2) + \gamma(k)$ in
the low angle regime of the SLC/LEP luminometers, where $\theta_{e^+},
\theta_{e^-} \in [25\ \mathrm{mrad}, 70\ \mathrm{mrad}]$ if $\theta_{e^+},
\theta_{e^-}$ are the CMS scattering angles of $e^+$, $e^-$ in the
$Z^0$ resonance energy regime.  The kinematics is illustrated in
Fig.~\ref{fig:process}.  It can be
shown~\cite{up-dn,kuraev-gatlinburg:1994} that, in this low angle
regime, graphs involving the exchange of more than one virtual photon
line between different fermion lines are suppressed in the
${\cal O}(\alpha^2)$ correction to the cross section.  Thus, we do not
need to calculate these graphs for the ${\cal O}(\alpha^2)$ corrections of
interest to us here.  Further, it can be shown that those terms in the
cross section involving interference of the radiation from the $e^+$
line with that from the $e^-$ line, in the low-angle Bhabha scattering
(LABH) regime, are severely suppressed as well --- this is the
so-called up-down interference suppression \cite{up-dn}.  Hence, we
will only need to calculate the graphs, as those shown in
Fig.\ \ref{fig:process}, where only one photon is exchanged between $e^+$
and $e^-$ lines and in which a virtual correction exists on one of
these lines. There are a total of 36 such graphs, excluding vacuum
polarization, and the ten electron-line emission graphs giving non-trivial 
results in our on-shell renormalization scheme are shown. The associated
positron-line emission graphs must be calculated as well. The $s$-channel 
exchange graphs will also be calculated for completeness and we will see 
that they are indeed negligible at the level of accuracy of interest to us
here.

For the actual calculation, we rely on two technical tools.  We
evaluate the general $\gamma$ emission amplitudes using the
formulation of the CALKUL \cite{CALKUL1,CALKUL2} methods given by Xu
{\em et al.}  \cite{magic}.  Thus, our massless fermion spinors
$|p,\pm\rangle$, of helicity $\pm$, are such that their spinor product
is
\begin{eqnarray}
\label{spr}
\langle p|p'\rangle_+ &\equiv& \langle p,-|p',+\rangle 
= (\mathbf{p}_x + i\mathbf{p}_y){\sqrt{p'_+}\over\sqrt{p_+}}
- (\mathbf{p'}_x + i\mathbf{p'}_y){\sqrt{p_+}\over\sqrt{p'_+}}\\
\langle p|p'\rangle_- &=& \langle p'|p\rangle_+^\ast \nonumber
\end{eqnarray}
for $p_+ = p^0 + \mathbf{p}_z$ in an obvious cartesian coordinate notation 
for the 3-momentum $\mathbf{p}$\ (i.e. 
$p^\mu=(p^0,p^1,p^2,p^3)=(p^0,\mathbf{p})=(p^0,p_x,p_y,p_z)$), 
and our corresponding photon polarization vectors are, for helicity $\rho$, 
\begin{equation}
\epsilon_\mu(k,h,\rho) = {\rho\over\sqrt{2}}
   {\langle h,-\rho|\gamma_\mu|k,-\rho\rangle\over\langle h|k\rangle_\rho}\ ,
\end{equation}
where $k$ is the photon momentum and $h$ is an auxiliary massless
4-vector.  As usual, $h$ may be chosen to simplify a given
gauge-invariant set of graphs\cite{magic}.  For a given helicity, three
of the ten graphs in Fig.\ \ref{fig:process} can be eliminated in this
manner, leaving seven to be evaluated.  The methods of Xu {\em et
al.}  are augmented for the evaluation of virtual corrections by the
algebraic manipulation program FORM\cite{FORM}, which we use to
evaluate our typical one-loop integrals via reduction to the
now-standard scalar integrals; here we use the realization of the
standard scalar integrals of Refs.\ \cite{Vermaseren,FF}, since they
were numerically stable enough for our applications.  It was with the
aid of these two computational techniques that we have calculated our
results for the ${\cal O}(\alpha)$ correction to the single bremsstrahlung
process $e^+ e^- \rightarrow e^+ e^- + \gamma$ at low SLC/LEP
luminosity regime angles.  Our results are presented in the next
section.


\section{Exact $\mathbf{{\cal O}(\alpha^2)}$ Results for 
$\mathbf{e^+ e^- \rightarrow e^+ e^- + \gamma}$}
\label{sec:calc}

In this section we present our results for the ${\cal O}(\alpha)$ correction
to the process $e^+ e^- \rightarrow e^+ e^- + \gamma$ at low angles. We
discuss first the $2\gamma$ bremsstrahlung effect.

For our $2\gamma$ bremsstrahlung correction, we use the results of three
of us (S.J., B.F.L.W, and S.A.Y.)  in Ref.\ \cite{twopho}, where the exact
result for $e^+ e^- \rightarrow e^+ e^- + 2\gamma$ was computed with the
methods of Xu {\em et al.} \cite{magic}. These results have been checked
in Ref.\ \cite{check}.

For the virtual correction to $e^+ e^- \rightarrow e^+ e^- + \gamma$,
we organize our results in terms of the amplitude for $\gamma$ emission
from the electron line and for emission from the positron line, neglecting
the so-called up-down interference terms~\cite{up-dn}, since these are known
to contribute negligibly at the level of our current precision of interest.

The electron-line emission amplitude with one virtual photon is
\begin{equation}
A_{(1)}^{e^-} = {ie^5\over 16\pi^2 t_p} 
\left( {\cal F}_{0} {\cal I}_0 + {\cal F}_{1} {\cal I}_1 + 
{\cal F}_{2} {\cal I}_2 \right)
\label{eq:amp}
\end{equation}
where
\begin{eqnarray}
{\cal I}_0 &=& 2\sqrt{2}\rho {\langle p_1, p_2\rangle_{-\rho}
 \left(\langle q_j, p_i \rangle_\rho\right)^2\over
 \langle q_1,k\rangle_{\rho} \langle q_2,k \rangle_{\rho}}\ ,\label{eq:I0}\\
{\cal I}_1 &=& 2\sqrt{2}\mu
{\langle q_{\hat{\jmath}},k\rangle_{-\rho}
 \langle p_2,q_j\rangle_{-\lambda} \langle q_j,p_1\rangle_{\lambda}\over
 \langle q_{\hat{\jmath}}, k \rangle_{\rho} 
 \langle q_1,q_2 \rangle_{-\rho}}\ ,\label{eq:I1}\\
{\cal I}_2 &=& 2\sqrt{2}\mu
{\langle q_{\hat{\jmath}},k\rangle_{-\rho} 
 \langle p_2,{q}_{\hat{\jmath}}\rangle_{-\lambda} 
 \langle {q}_{\hat{\jmath}},p_1\rangle_{\lambda}\over
 \langle q_{\hat{\jmath}}, k \rangle_{\rho} 
 \langle q_1,q_2 \rangle_{-\rho}}\ ,\label{eq:I2}
\end{eqnarray}
where the helicity-dependent indices $i, j, \widehat{\jmath}$ are given by
\begin{equation}
i = \cases{1\cr 2\cr}
\hbox{\ if $\rho = \pm\lambda$,} \qquad
(j,\widehat{\jmath}) = \cases{(1,2)\cr (2,1)\cr}
\hbox{\ if $\rho = \pm\mu$,}
\end{equation}
where $\lambda, \mu, \rho$ are the helicities of the positron,
electron, and photon.  
The function ${\cal I}_0$ is proportional to the electron line Born 
amplitude:
\begin{equation}
A^{\hbox{\scriptsize e${}^-$}}_{\mathrm{Born}} =
{i e^3\over t_p}\ {\cal I}_0\ .
\end{equation}

The form factors are 
\begin{eqnarray}\lefteqn{{\cal{F}}_0(\rho=\mu)= -8 - 8 m_e^2 C^{[0]}_{123} + 
    r_1(t_q - r_1)^{-1} - 2(t_q C_{124} + t_p C_{134}^{[0]})}\nonumber\\*
&-&\{t_q C_{124}\; - {r_1}C^{[r_1]}_{123}\; - (t_q + r_2)C^{[r_1]}_{134}\; 
    + (r_1 - r_2)C_{234}\; + {r_1}t_q D^{[r_1]}_{1234}\;\} \nonumber\\*
& &     \qquad\qquad\times\ t_q r_2^{-1}(r_1 - r_2)(t_q - r_1)^{-1} 
        \nonumber\\*
&+&\{t_q C_{124}\; + {r_2}C^{[-r_2]}_{123}\; - (t_q - r_1)C^{[-r_2]}_{134}\; 
    + (r_1 - r_2)C_{234}\; - {r_2}t_q D^{[-r_2]}_{1234}\;\}  \nonumber\\*
&+&6B_{12}\ +\ (B^{[r_1]}_{13}-B_{34})\;r_1(t_q-r_1)^{-1}\;
        \{1 - 3t_q (t_q + r_2)^{-1}\} \nonumber\\*
&-&6B_{34}\ +\ (B_{24}-B_{34})\;\{2t_q r_1(r_1 - r_2)^{-1}
        (t_q - r_1)^{-1} \} , \nonumber\\*
\label{eq:F0}
\end{eqnarray}
\begin{eqnarray}\lefteqn{{\cal{F}}_1(\rho=\mu)= 2t_q (r_1 - r_2)^{-1} 
        - t_q (t_q - r_1)^{-1}}\nonumber\\*
&+&\{t_q C_{124}\; - {r_1}C^{[r_1]}_{123}\; - (t_q  + r_2)C^{[r_1]}_{134}\; 
   + (r_1 - r_2)C_{234}\; + {r_1}t_q D^{[r_1]}_{1234}\;\} \nonumber\\*
& & \qquad\qquad\times\{t_p t_q r_2^{-2}(t_q - r_1)^{-1}(t_q - r_2) 
   +{\mbox{${1\over2}$}}\delta_{\rho,1} \} \nonumber\\*
&-&\{t_q C_{124}\; + {r_2}C^{[-r_2]}_{123}\; - (t_q - r_1)C^{[-r_2]}_{134}\; 
   + (r_1 - r_2)C_{234}\; - {r_2}t_q D^{[-r_2]}_{1234}\;\} \nonumber\\*
& &     \qquad\qquad\times\{{\mbox{${1\over2}$}}
    r_1^{-1}r_2\delta_{\rho,-1} \} \nonumber\\*
&+&(B^{[r_1]}_{13}-B_{34})\;t_p t_q (t_q - r_1)^{-1}\;\{2r_2^{-1} 
    -3(t_q + r_2)^{-1} \} \nonumber\\*
&+&2(B_{24}-B_{34})\;t_p t_q(r_1 - r_2)^{-1}\{(r_1 - r_2)^{-1}
    -t_q r_2^{-1}(t_q - r_1)^{-1}  \}, \nonumber\\*
\label{eq:F1}
\end{eqnarray}
\begin{eqnarray}\lefteqn{{\cal{F}}_2(\rho=\mu)= -2t_q(r_1 - r_2)^{-1} 
    + t_q(t_q + r_2)^{-1} }\nonumber\\*
&-&\{t_q C_{124}\; - {r_1}C^{[r_1]}_{123}\; - (t_q + r_2)C^{[r_1]}_{134}\; 
    + (r_1 - r_2)C_{234}\; + {r_1}t_q D^{[r_1]}_{1234}\;\} \nonumber\\*
& & \qquad\qquad\times\{t_p {t_q}r_2^{-2} 
    + {\mbox{${1\over2}$}}r_1r_2^{-1}\delta_{\rho,1} \} \nonumber\\*
&+&\{t_q C_{124}\; + {r_2}C^{[-r_2]}_{123}\; - (t_q - r_1)C^{[-r_2]}_{134}\; 
    + (r_1 - r_2)C_{234}\; - {r_2}t_q D^{[-r_2]}_{1234}\;\} \nonumber\\ *
    & & \qquad\qquad\times\{{\mbox{${1\over2}$}}\delta_{\rho,-1} \} \nonumber\\*
&+&(B_{34}-B^{[r_1]}_{13})\;t_p t_q(t_q+r_2)^{-1}\;\{2r_2^{-1} 
    + (t_q + r_2)^{-1} \} \nonumber\\*
&+&2(B_{24}-B_{34})\;t_p t_q(r_1-r_2)^{-1}\;\{r_2^{-1} 
    -(r_1 - r_2)^{-1} \}.\nonumber\\*
\label{eq:F2}
\end{eqnarray}

The opposite helicity cases may be obtained from the above results using
the substitutions (for $i = 0,1,2$)
\begin{equation}
{\cal F}_i(\rho=-\mu, r_1,r_2) = 
{\cal F}_i(\rho=\mu, -r_2, -r_1)\ .
\end{equation}

The scalar integrals $B$, $C$, $D$ are
defined in Refs.  \cite{Vermaseren} (we evaluate them using algorithms
from Refs.  \cite{Vermaseren}) and in Appendix \ref{app:integrals}.  The
kinematic variables $s, s', t_p, t_q, r_i$ are defined as
\begin{eqnarray}
s &=& (p_1+q_1)^2,\qquad s'\ =\ (p_2+q_2)^2,\nonumber\\
t_p &=& (p_1-p_2)^2,\qquad t_q\ =\ (q_1 - q_2)^2, \\
& &\qquad\qquad r_i\ =\ 2q_i\cdot k. \nonumber
\label{eq:kinconv}
\end{eqnarray}

We have found that at low angles in the SLC/LEP energy regime, the 
${\cal F}_0{\cal I}_0$ terms in (\ref{eq:amp}) are often a good 
approximation to the entire result, and that these terms are in turn well 
approximated by the simple expression
\begin{eqnarray}
\label{eq:approx}
A^{e^-}_{(1)\ \mathrm{approx}} &=& {ie^5\over 16\pi^2 t_p}
  \ {\cal F}_{0}^{\ \mathrm{approx}}\quad{\cal I}_0\ ,  \\[2ex]
{\cal F}_{0}^{\ \mathrm{approx}} &=& 
  -8 - 8 m_e^2 C^{[0]}_{123} - 2(t_q C_{124} + t_p C^{[0]}_{134}) 
  + 6 B_{12} - 6 B_{34} 
\nonumber\\
   &=& -\;8 + \frac{2\pi^2}{3} - \ln^2 \frac{|t_p|}{m_e\!^2} - \ln^2
     \frac{|t_q|}{m_e\!^2} + 6 \ln \frac{|t_p|}{m_e\!^2} \nonumber \\
     & & +\ 4 \ln \frac{|t_p|}{m_e\!^2} \ln \frac{m_\gamma}{m_e} 
     + 4 \ln \frac{|t_q|}{m_e\!^2} \ln \frac{m_\gamma}{m_e} 
     - 8 \ln \frac{m_\gamma}{m_e} .\nonumber\\
\end{eqnarray}
This approximation has been compared to the complete result in
detail and we have found it to be within $~10\%$ of the result (\ref{eq:amp})
throughout most of the final particle phase space; such
comparisons will be presented
in more detail elsewhere\cite{semi-paper}.

The analog of (\ref{eq:amp}) for positron line emission is obtained by
crossing: 
\begin{equation}
A_{(1)}^{e^+} \ = \ A_{(1)}^{e^-}\ (p_1 \leftrightarrow -q_2', 
      q_1 \leftrightarrow -p_2,\lambda \leftrightarrow -\mu).
\end{equation}
The differential cross section associated with (\ref{eq:amp}) is the usual
\begin{equation}
{d\sigma^{{\cal O}(\alpha^2)}\over d\Omega kdkd\Omega_k} =
{({p_2}^0)^2\over (4\pi)^5 ss'} \quad
\sum_{\lambda,\mu,\rho}
\mathrm{Re}\ \left(A^{e^+}_{(1)} + A^{e^-}_{(1)}\right)
\left(A^{e^+}_{\mathrm{Born}} + A^{e^-}_{\mathrm{Born}}\right)^{\ast} ,
\label{eq:dcs}
\end{equation}
where $\Omega$ is the outgoing positron solid angle in the lab frame,
$\Omega_k$ is the photon solid angle, and the up-down interference terms 
can be neglected here.

We have used massless fermion spinors to calculate (\ref{eq:amp})
-- (\ref{eq:dcs}).  The relevant mass effects can be restored by the
standard methods already published in Ref.\ \cite{CALKUL2}.  We have
done that, and it amounts to adding an $m_e\!^2$-correction term for
each external emission line in Fig.\ \ref{fig:process}.  The
corresponding corrections are given in the Appendix \ref{app:mass} for
completeness.  In this way, we have arrived at an exact ${\cal O}(\alpha)$
correction to the process $e^+ e^- \rightarrow e^+ e^- + \gamma$ with
all relevant mass effects taken explicitly into account.


\section{Results and Checks}
\label{sec:results}

Our results (\ref{eq:amp}) -- (\ref{eq:dcs}) are readily introduced
into the Monte Carlo program BHLUMI \cite{th6118:1991,th-9538:1995} of
two of us.  This will be presented elsewhere\cite{semi-paper}.  Here,
we wish to discuss some numerical checks we have made on these
results.

\begin{figure}
\begin{center}
\epsfig{file=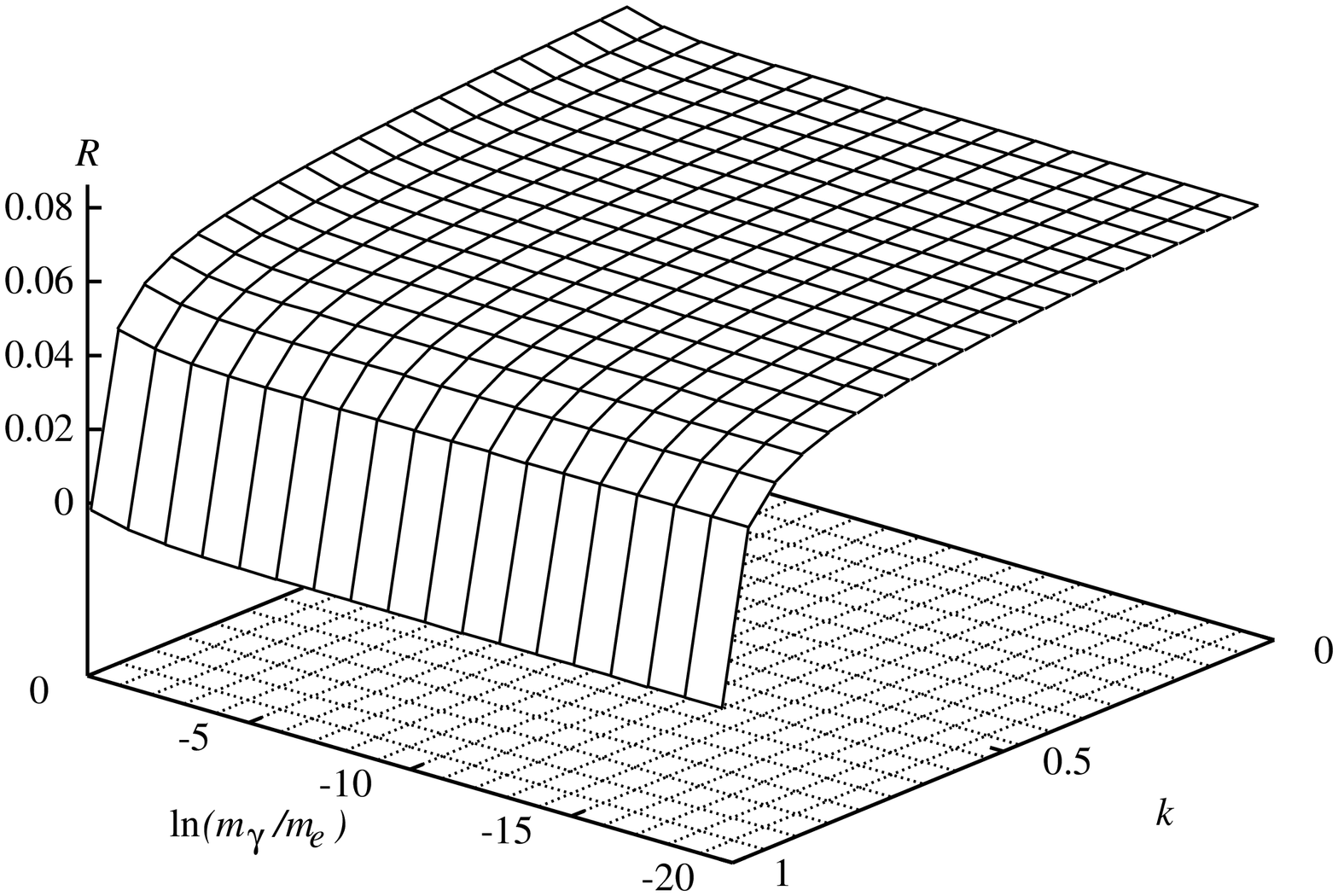,angle=270,width=5in}
\end{center}
\caption{$\ln m_\gamma$ dependence of the ratio 
$R = {d\sigma^{{\cal O}(\alpha^2)}_{\mathrm{IR-reg}}\over 
      d\Omega kdkd\Omega_k} {\Big/} 
    {d\sigma^{{\cal O(\alpha)}}\over d\Omega kdkd\Omega_k}$
where the numerator
is defined in (\protect\ref{eq:IR-reg}) and the denominator
is the ${\cal O}(\alpha)$ single bremsstrahlung cross-section.
We plot $R$ as a function of $k = 1 - s'/s$ and $\ln(m_\gamma/m_e)$ for 
positron polar angles $(\theta,\phi) = (2^\circ, 0^\circ)$ and photon polar 
angles $(\theta,\phi) = (1.5^\circ, 10^\circ)$ with respect to the positron 
beam axis. This shows that $\ln m_\gamma$ has cancelled out when $m_\gamma 
\ll m_e$, so that the result is independent of the infrared regulator.}
\label{fig:regulator}
\end{figure}

For our checks, we define an IR-regular differential cross section
by subtracting the virtual infrared contribution as given by the YFS 
theory\cite{yfs:1961}:
\begin{equation}
{d\sigma^{{\cal O}(\alpha^2)}_{\mathrm{IR-reg}}\over d\Omega kdkd\Omega_k} 
\ =\ {d\sigma^{{\cal O}(\alpha^2)}\over d\Omega kdkd\Omega_k}\ -
   \ 2\alpha (\mathop{\mathrm{Re}} {B}_{\mathrm{YFS}})\;
   {d\sigma^{{\cal O}(\alpha)}\over d\Omega kdkd\Omega_k}\ ,
\label{eq:IR-reg}
\end{equation}
with $B_{\mathrm{YFS}}$ defined as in Ref.\ \cite{JW:1988}. We have verified
that our infrared limit $k\rightarrow 0$ is correct and in agreement with
the YFS theory\cite{yfs:1961} of that limit. This agreement with YFS
theory is further illustrated in Fig.~\ref{fig:regulator}, where we show 
that (\ref{eq:IR-reg}) is independent of our photon regulator mass $m_\gamma$.

Further, looking into the results in Fig. 2, we see, in addition to
their independence of $\ln m_\gamma$, that the size of our
${\cal O}(\alpha^2)$ virtual correction is $\sim 0.05$ of the respective real
${\cal O}(\alpha)$ correction itself, which is consistent with the naive
LL power counting expectations, since ${\alpha\over\pi}L\sim 0.05$ here.
Illustrations of the actual size of the subleading $\alpha^2L$ part of the
exact result (\ref{eq:amp}) in the ALEPH SICAL-type acceptance in connection 
with BHLUMI4.xx will appear elsewhere\cite{semi-paper}.

For completeness, we note that we have also evaluated the exact result
for the s-channel exchange contribution
to the ${\cal O}(\alpha^2)$ correction to the cross sections
evaluated here. We find that it is below $0.04$ of the 
${\cal O}(\alpha^2)$ correction itself. This shows that such exchanges
are indeed negligible at the level of precision of interest to us here
in the evaluation of our respective ${\cal O}(\alpha^2)$ effects.

Thus, the important exact ${\cal O}(\alpha)$ correction to $e^+ e^-
\rightarrow e^+ e^- + \gamma$ has now been calculated and it has been
verified that it has the correct IR limit as well as a size consistent
with the naive power counting expectations.
The
implementation of this exact result into the Monte Carlo event
generator BHLUMI4.xx \cite{lep1-94a,th-9538:1995} for LABH is now in
progress and will appear elsewhere\cite{semi-paper}.


\section{Conclusion}
\label{sec:final} 

In this paper, we have presented the first complete exact ${\cal O}(\alpha)$ 
correction to $e^+ e^- \rightarrow e^+ e^- + \gamma$ at low angles in the 
SLC/LEP energy regime.  We have checked the infrared limit of our work 
against known expectations.  We have evaluated our results in the ALEPH 
SICAL type acceptance and found that the respective correction is
consistent in size with the naive power counting expectations for its LL
content.
Implementation of our results into the Monte Carlo event
generator BHLUMI \cite{th-9538:1995} for LABH is in progress and will
appear elsewhere\cite{semi-paper}.

We should note that the authors in Ref.\ \cite{kuraev-gatlinburg:1994}
have also given pioneering
results for this ${\cal O}(\alpha)$ correction to $e^+ e^-
\rightarrow e^+ e^- + \gamma$ at low angles.  We have not
compared our results with theirs however because they have not
provided us with a final version of their fully differential results
to date. We await their final published expressions with which we may 
compare.

In summary, the dominant missing contribution to the ${\cal O}(\alpha^2)$
brems\-strah\-lung effect in the theoretical prediction for the SLC/LEP
luminosity process $e^+ e^- \rightarrow e^+ e^-$, the ${\cal O}(\alpha^2
L)$ correction, has now been calculated exactly.  We look forward with
excitement to the application of this correction to reach a new level
of precision on the respective theoretical prediction, via BHLUMI4.xx
\cite{th-9538:1995}, of this luminosity.  Such applications will
appear elsewhere\cite{semi-paper}.


\newpage
\section*{Acknowledgments}
Two of the authors (S.J.\ and B.F.L.W.) would like to thank 
Profs.\ G.\ Altarelli and G.\ Veneziano of the CERN TH Div.\ and the ALEPH 
Collaboration for their support and hospitality while this work was completed. 
B.F.L.W.\ would like to thank Prof.\ C.\ Prescott of Group A at SLAC for his 
kind hospitality while this work was in its developmental stages.


\appendix
\section*{Appendices}
\section{Scalar Integrals}
\label{app:integrals}

In this appendix, we define the scalar integrals used in the formulas in
the text. We have, from Ref.\ \cite{Vermaseren}, with the kinematic conventions
(\ref{eq:kinconv}), 
\begin{eqnarray}
B_{12} &=& B(m_e\!^2; m_\gamma, m_e) \\
B_{13}^{[r]} &=& B(m_e\!^2 - r; m_\gamma, m_e) \\
B_{24} &=& B(t_q;m_e,m_e) \\
B_{34} &=& B(t_p;m_e,m_e) \\
C_{123}^{[r]} &=& C(m_e\!^2, m_{\gamma}\!^2, m_e\!^2 - r; 
  m_\gamma, m_e, m_e) \\
C_{124} &=& C(m_e\!^2, t_q, m_e\!^2; m_\gamma, m_e, m_e) \\
C_{134}^{[r]} &=& C(m_e\!^2 - r, t_p, m_e\!^2; m_\gamma, m_e, m_e) \\
C_{234} &=& C(m_{\gamma}\!^2, t_p, t_q; m_e, m_e, m_e) \\
D_{1234}^{[r]} &=& 
  D(m_e\!^2, m_{\gamma}\!^2, t_p, m_e\!^2, m_e\!^2 - r, t_q; 
  m_\gamma, m_e, m_e, m_e)
\end{eqnarray}
where we have defined the basic scalar integrals as
\begin{eqnarray}
& &B(p^2,m_1,m_2) = {(2\pi\mu)^{4-D}\over\pi^2 i}\int 
    {d^D q\over(q^2 - m_1\!^2 + i\epsilon) ((q+p)^2 - m_2\!^2 + i\epsilon)}\ ,
\nonumber\\
& & 
\label{eq:Bdef}\\
& &C(p_1^2, p_2^2, (p_1+p_2)^2; m_1, m_2, m_3) =\nonumber\\*
& &  {1\over\pi^2 i}\int 
    {d^4 q\over(q^2 - m_1\!^2 + i\epsilon)((q+p_1)^2 - m_2\!^2 + i\epsilon)
    ((q+p_1+p_2)^2 - m_3\!^2 + i\epsilon)}\ ,\nonumber\\
& & 
\label{eq:Cdef}\\
& &D(p_1^2, p_2^2, p_3^2, (p_1+p_2+p_3)^2, (p_1+p_2)^2, (p_2+p_3)^2; 
    m_1, m_2, m_3, m_4) =\nonumber\\*
& &{1\over\pi^2 i}\int d^4 q\ {1\over(q^2 - m_1\!^2 + i\epsilon)
           ((q+p_1)^2 - m_2\!^2 + i\epsilon)}\nonumber\\*
& & \qquad\times {1\over((q+p_1+p_2)^2 - m_3\!^2 + i\epsilon)
           ((q+p_1+p_2+p_3)^2 - m_4\!^2 + i\epsilon)}\ .\nonumber\\
& & 
\label{eq:Ddef}
\end{eqnarray}

The $B$ integral is defined using dimensional regularization,
and the $C$ and $D$ integrals are UV-finite. The final amplitude
is independent of the mass scale $\mu$ in the definition of $B$.


\section{Mass Corrections}
\label{app:mass}
In this appendix, we give our mass correction for the cross section in 
(\ref{eq:dcs}). Specifically, following Ref.~\cite{CALKUL2}, we find that 
the mass correction to (\ref{eq:dcs}) is 
\begin{equation}
{d\;\Delta\sigma_{m}\over d\Omega kdkd\Omega_k} =
{(p_2^{0})^2\over 2^9 \pi^5 ss'} |A_m|^2 , 
\end{equation}
where we have defined
\begin{equation}
|A_m|^2 \ =\  - \  \frac{e^2m^2}{(qk)^2} f_0(q-k,p_i) .
\label{eq:genmasscorr}
\end{equation}
Here, the photon is radiated nearly parallel to $q$, and $f_0$ denotes the 
non-radiative cross section, summed over all polarizations, with the original 
$q$ replaced by $q-k$. In the case of Bhabha scattering the Born cross 
section is proportional to the following invariant summed matrix element 
squared: 
\begin{equation}
f_B\!^{e^+\!+e^-} \ =\  {2e^4\over t^2}(s^2+u^2).
\end{equation}
The complete non-radiative cross section for the O($\alpha^2$) single 
bremsstrahlung mass corrections is then proportional to
\begin{equation}
f_0\!^{e^+\!+e^-} \ =\  (1+ \frac{e^2}{4\pi^2} {\cal F})f_B\!^{e^+\!+e^-} 
\end{equation}
with 
\begin{equation}
{\cal F}(t) = 
\ =\   2\left(\ln \frac{|t|}{m_e\!^2} -1\right) .
\label{eq:softff}
\end{equation}
From (\ref{eq:genmasscorr}) it follows that, when summed over all fermion 
legs, the finite mass terms for the O($\alpha^2$) single bremsstrahlung 
corrections are given by
\begin{eqnarray}
&&|A_{m_e}\!\!\!^{e^+\!+e^-}|^2 = \nonumber\\*
&&-\; {2e^6m_e\!^2\over(q_1 k)^2}\; 
\left[ 1 + \frac{e^2}{4\pi^2} {\cal F}(-2q_1 q_2+2q_2 k) \right]\; 
\ {(p_1 q_1-p_1 k)^2 + (q_1 p_2-p_2 k)^2\over(q_1 q_2-q_2 k)^2} 
\nonumber \\*
&&-\; {2e^6m_e\!^2\over(q_2 k)^2}\; \left[ 1 + \frac{e^2}{4\pi^2} 
{\cal F}(-2q_1 q_2-2q_1 k) \right] \;
\ {(p_2 q_2+p_2 k)^2 + (p_1 q_2+p_1 k)^2\over(q_1 q_2+q_1 k)^2} 
\nonumber \\*
&&- \; {2e^6m_e\!^2\over(p_1 k)^2}\; \left[ 1 + \frac{e^2}{4\pi^2} 
{\cal F}(-2p_1 p_2+2p_2 k) \right] \; 
\ {(p_1 q_1-q_1 k)^2 + (p_1 q_2-q_2 k)^2\over(p_1 p_2-p_2 k)^2} 
\nonumber \\*
&&- \; {2e^6m_e\!^2\over(p_2k)^2}\; \left[ 1 + \frac{e^2}{4\pi^2} 
{\cal F}(-2p_1 p_2-2p_1k) \right] \; 
\ {(p_2 q_2+q_2 k)^2 + (q_1 p_2+q_1 k)^2\over(p_1 p_2+p_1 k)^2} {\ }_{.}
\nonumber\\*
\end{eqnarray}

This completes our appendices.


\bibliographystyle{unsrt}

\begin{thebibliography}{10}

\bibitem{lumin:1995}
{See for example, B.~Pietrzyk},
\newblock In B.F.L. Ward, ed., {\em {Proc.\ Tennessee Intl.\ Symp.\ on
  Radiative Corrections: Status and Outlook}} (World Scientific, Singapore,
  1995) p.\ 138.

\bibitem{th6118:1991}
{S. Jadach, E. Richter-W\c as, B.F.L. Ward and Z. W\c as},
\newblock {\em Phys. Lett.} {\bf B}268 (1991) 253.

\bibitem{glasgow}
{B.F.L.\ Ward, S.\ Jadach, E.\ Richter-W\c{a}s and Z.\ W\c{a}s},
\newblock {\em Precision Calculation of the Small Angle Bhabha Cross 
Section},
\newblock {presented by B.F.L.\ Ward at the Rochester Conference, 
Glasgow, UK}, July, 1994.

\bibitem{th-9538:1995}
{S.~Jadach, E.~Richter-W\c as, B.F.L.~Ward and Z.~W\c as},
\newblock {{preprint CERN-TH-95-38, in press at {\em Phys.\ Lett.}\ {\bf B} 
(1995)}}.

\bibitem{gatlin:1994}
{S.~Jadach, M.~Melles, W.~P\l aczek, E.~Richter-W\c as, M.~Skrzypek, 
B.F.L.~Ward, Z.~W\c as and S.~Yost},
\newblock In B.F.L. Ward, ed., {\em {Proc.\ Tennessee Intl.\ Symp.\ on
  Radiative Corrections: Status and Outlook}} (World
  Scientific, Singapore, 1995) p.\ 153.

\bibitem{semi-paper}
S.~Jadach et~al,
\newblock in preparation.

\bibitem{zhang}
{S.\ Jadach, B.F.L.\ Ward, E. Richter-W\c{a}s and H.\ Zhang},
\newblock {\em Phys.\ Rev.} {\bf D42} (1990) 2997.

\bibitem{twopho}
{S.\ Jadach, B.F.L.\ Ward and S.A.\ Yost},
\newblock {\em Phys.\ Rev.} {\bf D47} (1993) 2682.

\bibitem{lep1-94a}
{S.\ Jadach, M.\ Melles, W.\ P\l aczek, E.\ Richter-W\c{a}s, M.\ Skrzypek,
  B.F.L.\ Ward, Z.\ W\c{a}s and S.\ Yost },
\newblock {\em Higher-Order Radiative Corrections to Bhabha Scattering 
at Low Angles: The YFS Monte Carlo Approach}.
\newblock {CERN Yellow Report 93-03, Chapter 12}, 1994.

\bibitem{kuraev-gatlinburg:1994}
{V.~Fadin, E.~Kuraev, L.~Lipatov, N.~Merenkov and L.~Trentadue},
\newblock In B.F.L. Ward, ed., {\em {Proc.\ Tennessee Intl.\ Symp.\ on
  Radiative Corrections: Status and Outlook}} (World Scientific, Singapore,
  1995) p.\ 168.

\bibitem{partial}
{F.\ A.\ Berends, W.\ L.\ van Neerven and G.J.H.\ Burgers},
\newblock {\em Nucl.\ Phys.} {\bf B297} (1989) 429,
\newblock {{\it ibid.} {\bf B304} (1988) 921}.

\bibitem{FORM}
{J.A.M.\ Vermaseren}.
\newblock {The symbolic manipulation program FORM, Version 1.0},
\newblock {available via anonymous FTP from nikhefh.nikhef.nl}.

\bibitem{up-dn}
{S.\ Jadach, E.\ Richter-W\c as, B.F.L.\ Ward and Z.\ W\c as},
\newblock {\em Phys.\ Lett.} {\bf B253} (1991) 469.

\bibitem{CALKUL1}
{F.A.\ Berends, P.\ De Causmaecker, R.\ Gastmans, R.\ Kleiss, 
W.\ Troost and T.T.\ Wu},
\newblock {\em Nucl.\ Phys.} {\bf B239} (1984) {382, 395}.

\bibitem{CALKUL2}
{F.A.\ Berends, P.\ De Causmaecker, R.\ Gastmans, R.\ Kleiss, 
W.\ Troost and T.T.\ Wu},
\newblock {\em Nucl. Phys.} {\bf B264} (1986) {243, 265}.

\bibitem{magic}
{Z.\ Xu, D.-H.\ Zhang and L.\ Chang},
\newblock {\em Nucl.\ Phys.} {\bf B291} (1987) 392.

\bibitem{Vermaseren}
{G.J.\ van Oldenborgh and J.A.M.\ Vermaseren},
\newblock {\em Zeit.\ Phys.} {\bf C46} (1990) 425.

\bibitem{FF}
{G.J.\ van Oldenborgh},
\newblock {\em FF, a package to evaluate one-loop Feyman diagrams},
\newblock NIKHEF-H/90-15, 1990.

\bibitem{check}
{{O.\ Adriani \it et al.}},
\newblock {\em Phys. Lett.} {\bf B295} (1992) 337.
\newblock {E.\ Richter-W\c as, unpublished; K.\ Riles, unpublished.}

\bibitem{yfs:1961}
D.R. Yennie, S.~Frautschi, and H.~Suura,
\newblock {\em Ann. Phys. (NY)} {\bf 13} (1961) 379.

\bibitem{Kleiss}
{F.A.\ Berends and R.\ Kleiss},
\newblock {\em Nucl.\ Phys.} {\bf B228} (1983) 537.

\bibitem{JW:1988}
{S.\ Jadach and B.F.L.~Ward}, 
\newblock {\em Phys.\ Rev.} {\bf D38} (1988) 2897.

\bibitem{Bohm}
{M.\ B\"ohm, H.\ Spiesberger and W.\ Hollik},
\newblock {\em Fort.\ Phys.} {\bf 34} (1986) 11.

\end{thebibliography}

\end{document}